\documentclass[fleqn,12pt,twoside]{article}
\usepackage{espcrc1}
\usepackage{epsfig}

\usepackage{graphicx}
\usepackage[figuresright]{rotating}

\newcommand{\AmS}{{\protect\the\textfont2
  A\kern-.1667em\lower.5ex\hbox{M}\kern-.125emS}}

\title{Confinement: Results and Perspectives}

\author{Adriano Di Giacomo\address[MCSD]{Universit\`a di Pisa and INFN, \\ 
        Dipartimento di Fisica, Via Buonarroti 2, 56125 Pisa, Italy}%
        \thanks{Partially supported by MURST and by EC, FMRX-CT97-0122},
        }

\def\be{\begin{equation}}
\def\ee{\end{equation}}  
\begin{document}

\maketitle

\begin{abstract}
An updated review is presented of lattice investigations of colour confinement.
\end{abstract}

\section{Introduction}
The existing experimental evidence for confinement is based on the search of
fractionally charged particles (quarks). Such particles have not been observed,
and upper limits have been established which are typically $10^{-15}$ times
smaller than what one would expect in the absence of confinement\cite{1}.
The natural interpretation of that is that quarks do not exist as free
particles,and that this absolute property is based on a symmetry\cite{2}.

No direct evidence exists instead for confinement of other colored particles with
zero or integer charge, e.g. gluons. The question is then: what symmetry of QCD
prevents the observation of free quarks and possibly of other colored particles?

A deconfinement transition is observed in numerical simulations of quenched QCD on
the lattice\cite{3}.

The theory at finite temperature $T$ is described by Euclidean Feynman integral
extending from time 0 to time $1/T$, with periodic boundary conditions in time for
bosons, antiperiodic for fermions. On a lattice $N_T\times N_S^3$ this means $T=1/a
N_T$. $N_T$ is the time extension, $N_S$ the space extension of the lattice, and
$N_S \gg N_T$.

In physical units $a$ depends on the bare coupling constant $g$, at sufficiently
small $g$, as 
\begin{equation}
a = \frac{1}{\Lambda_L} \exp(-b_0\beta)\qquad \beta = \frac{2N}{g^2}
\label{eq1}\end{equation}
Eq.(\ref{eq1}) stems from renormalization group and $-b_0$ is the coefficient of
the lowest term in the perturbative expansion of the beta function. Because of
asymptotic freedom $b_0 > 0$ and
$
T = \frac{\Lambda_L}{N_T}\exp(b_0\beta)$.
Low temperature corresponds to strong coupling regime (disorder), high temperature
to weak coupling (order). The
symmetry responsible for confinement is a symmetry of the disordered phase.

The way to observe confinement is to look at the spatial correlator of the Polyakov
lines
\be
G(\vec x-\vec y) \equiv \langle L(\vec x) L(\vec y)\rangle\label{eq3}
\ee
The Polyakov line
$
L(\vec x) = {\cal P}\exp\left[ i \int_0^{1/T} A_0(\vec x,\tau)\,d\tau\right]
$
is the parallel transport through the lattice along the time axis. The static
potential $V(\vec z)$ between a $q\,\bar q$ pair at distance $|\vec z|$ is given by
\be
V(\vec z) = \frac{d}{d|\vec z|} \ln G(\vec z)\label{eq5}
\ee
one has by cluster property
\be
G(\vec z) \mathop\simeq_{|\vec z|\to\infty} C\exp(-\sigma |\vec z|) +
|\langle L\rangle|^2\label{eq6}
\ee
Lattice data show that a temperature $T_C$ exists, such that at $T< T_C$ 
$|\langle L\rangle|=0$ so that
\be
V(\vec z) \mathop\simeq_{|\vec z|\to\infty} \sigma|\vec z|
\qquad \sigma = \hbox{string tension}
\label{eq7}
\ee
a linear potential which implies confinement.

At $T>T_C$ 
$|\langle L\rangle|\neq 0$
and $
V(\vec z) \mathop\simeq_{|\vec z|\to\infty} 0 $.

The resulting indication is that $|\langle L\rangle|$ is an order parameter and
$\sigma$ a disorder parameter for the deconfining transition.

For $SU(2)$ the transition is 2nd order and belongs to the universality class of
the 3d Ising model. For $SU(3)$ it is weak first order.

The underlying symmetry is $Z_N$, the centre of the gauge group. In full
QCD,however, $Z_N$ is explicitely broken by the very presence of quarks, and also
$\sigma$ does not exist due to the so called string breaking, i.e. to formation of
$q\,\bar q$ pairs when trying to pull $q\,\bar q$ apart. A chiral phase transition
exists in this case, from the Goldstone phase to the Wigner phase and the order
parameter is $\langle \bar\psi\psi\rangle$, the chiral condensate. The connection
of chiral transition to deconfinement is not clear.

This situation is unsatisfactory, especially in view of the philosophy of $N_C\to
\infty$. According to it the basic structure of the theory is determined at
$N_c=\infty$, corrections in $1/N_C$ being small and convergent. Quark loops are
non leading in this expansion. Therefore a good order (disorder) parameter should
not depend on the presence of quarks.

\section{Duality.}
To understand the symmetry of a disordered phase the concept of duality is
fundamental\cite{4}. Duality applies to $d-$dimensional systems, having non local
excitations with non trivial topology in $d-1$ dimensions.

Such systems admit two complementary descriptions.

A direct description interms of local fields $\Phi$, in which topological
excitations $\mu$ are non local. The direct description is convenient in the weak
coupling regime (ordered phase).

A dual description, in which the topological excitations $\mu$ are local operators
and the fields $\Phi$ non local excitations. The dual description is convenient in
the strong coupling regime  $g\gg 1$ (disordered phase). The effective coupling of
the dual $g_D$ is related to $g$ as $g_D\sim 1/g \ll 1$. Duality maps the strong
coupling regime of the direct description into the weak coupling of the dual.

Duality applies to a number of statistical systems, (3d XY model\cite{5}, 3d
Heisenberg model\cite{6}\ldots)the prototype being the 2d Ising model\cite{7,8}.
Gauge theories are additional examples (compact $U(1)$ gauge theory\cite{9,10},
$N=2$ SUSY QCD\cite{11}), and so are M-string theories\cite{12}. We will discuss
how it works for QCD.

The question is: what are the dual excitations of QCD, and what their quantum
numbers, which determine the symmetry of the confining vacuum.

Two main ideas exist:
\begin{itemize}
\item[a)]Monopoles. Their condensation produces dual superconductivity of the
vacuum and confinement of electric charges via dual Meissner effect\cite{13}.
\item[b)]Vortices\cite{2}. A vortice is a  defect associated to a closed line
$C$. If $B(C)$ is the operator which creates a vortex on $C$ and $W(C')$ the
operator which creates a Wilson loop on the contour $C'$, they obey the algebra
\be
B(C) W(C') = W(C') B(C) \exp\left[ i \frac{n_{CC'}}{N} 2\pi\right]
\label{eq9}
\ee
where $n_{CC'}$ is the winding number of the curves $C$ and $C'$.

>From the above algebra it follows\cite{2} that whenever $\langle W(C')\rangle$
obeys the area law $\langle B(C)\rangle$ obeys the perimeter law, and viceversa if
$\langle B(C)\rangle$ obeys the area law, $\langle W(C')\rangle$ obeys the
perimeter law. Some people speak of condensation of vortices: however the fact
that $\langle B(C)\rangle\neq 0$ has no special meaning for the symmetry, in the
same way as the fact that $\langle W(C')\rangle\neq 0$. What matters is the area
or perimeter law. This can be synthetized\cite{14} in terms of the Polyakov line $\langle
L\rangle$ and the dual Polyakov line (t'Hooft line) $\langle L'\rangle$.
Area law for $\langle W\rangle$ means $\langle L\rangle = 0$, perimeter law 
$\langle L\rangle\neq 0$. In the same way for $\langle B\rangle$ area law means
$\langle L'\rangle =0$, perimeter law $\langle L'\rangle\neq 0$.

So for $T< T_C$ one expects $\langle L\rangle = 0$, $\langle L'\rangle \neq 0$;
for $T > T_C$ $\langle L'\rangle = 0$, $\langle L\rangle \neq 0$.
$\langle L'\rangle$ is a disorder parameter, $\langle L\rangle$ an order parameter
for confinement.

In the presence of quarks however $B(C)$ is not only dependent on the line $C$,
but can only be defined on surfaces having $C$ as a contour, and depends on the
choice of the surface\cite{15}.
\end{itemize}
\section{Defining monopoles: the abelian projection.}
A conserved magnetic current $j_\mu^M(x)$ can be associated to any operator
$\vec\varphi(x)$ in the adjoint representation. We shall consider for simplicity the
case $SU(2)$: all what follows can be easily extended to generic $SU(N)$. 

Let us start by defining $\hat\varphi(x) = \vec\varphi(x)/|\vec\varphi(x)|$, a
direction in colour space. $\hat \varphi$ is defined everywhere except at sites
where $\vec \varphi(x)=0$. According to ref.(\cite{16}) we can define a singlet,
gauge invariant field strength
\be
F_{\mu\nu} = \hat\varphi \vec G_{\mu\nu} - \frac{1}{g}\hat\varphi\left(D_\mu
\hat\varphi\wedge D_\nu\hat\varphi)\right) =
\hat\varphi(\partial_\mu \vec A_\nu - \partial_\nu\vec A_\mu)
-\frac{1}{g}\hat\varphi(\partial_\mu\hat\varphi\wedge\partial_\nu\hat\varphi)
\label{eq10}
\ee
The two terms are separately gauge invariant, but the coefficients are chosen in
such a way that bilinear terms in $A_\mu A_\nu$ and $A_\mu\partial_\nu\hat\varphi$
cancel. 
Going to a gauge $\hat\varphi = const$ (eg. $\varphi(x) = (0,0,1)$) ({\em abelian
projection}) the last term in eq.(\ref{eq10}) becomes zero and $F_{\mu\nu}$
reduces to an abelian field\cite{13}
$
F_{\mu\nu} = \partial_\mu A^3_\nu - \partial_\nu A^3_\mu
$
The abelian projection is non singular except at zeros of $\vec\varphi(x)$ where,
after the projection, $U(1)$ Dirac monopoles appear.

Defining the dual tensor
$
F^*_{\mu\nu} = \frac{1}{2}\varepsilon_{\mu\nu\rho\sigma} F^{\rho\sigma}
$
and
$
J^M_\mu = \partial^\nu F^*_{\nu\mu}$,
because of the antisymmetry of $F^*_{\mu\nu}$, $J^M_\mu$ is conserved
\be
\partial^\mu J^M_\mu = 0\label{eq15}
\ee
In non compact formulations $J^M_\mu=0$ (Bianchi identities); in compact
formulations like lattices it can be nonzero\cite{17}.

The magnetic $U(1)$ symmetry implied by eq.(\ref{eq15}) can be 
\begin{itemize}
\item[a)] Wigner. The magnetic charge $Q$ is defined as an hermitian operator and
the Hilbert space splits into superselected sectors of definite magnetic charge.
In this case the vacuum expectation value $\langle\mu\rangle$ of any operator $\mu$
carrying magnetic charge vanishes
$
\langle\mu\rangle = 0$
\item[b)] Higgs broken. Then at least one magnetically charged operator $\mu$
exists, such that $\langle\mu\rangle\neq 0$. The free energy density (effective
lagrangean) can then be written by symmetry and dimensional arguments in terms of
$\langle\mu\rangle$ and $\tilde A_\mu$, the dual vector potential, as a Higgs
Lagrangean\cite{18} and dual superconductivity follows.
\end{itemize}
In conclusion, in non abelian gauge theories, a conserved magnetic current exists,
associated to any choice of the field $\vec\varphi$. $\vec\varphi$ is arbitrary,
except for the fact that it belongs to the adjoint representation. For $SU(N)$ the
construction provides $N-1$ conserved currents.

The idea\cite{19,20} is then that the $U(1)$ symmetry is Higgs broken in the
confined phase (monopoles condense in the vacuum, electric charges are confined by
dual Meissner effect), becomes Wigner at the deconfining transition and stays such
in the deconfined phase.

The dependence on the choice of $\vec \varphi$ is not a priori known.

This pattern is detectable by constructing a disorder parameter
$\langle\mu\rangle$, the v.e.v. of a magnetically charged operator\cite{19,20}. Then
$\langle\mu\rangle\neq0$, $T< T_C$ 
\[ \langle\mu\rangle \mathop\simeq_{T\to T_C^-} \left(1 -
\frac{T}{T_C}\right)^\delta\qquad \langle\mu\rangle = 0\quad T> T_C\]
\section{Construction of the disorder parameter $\langle\mu\rangle$.}
Expressing $\mu$ in terms of the fields $\varphi$ of the direct description is an
explicit realization of the duality transformation. The general idea is to shift
the fields by the classical, topological configuration\cite{21} which is created.
In the Scr\"odinger representation $|\varphi(x)\rangle$, this is done in terms of
the conjugate momentum operators
$\Pi(x)$, defined by the canonical commutation relations
\be
\left[\varphi(x_0,\vec x),\Pi(x_0,\vec y)\right] = i\,\delta^3(\vec x - \vec y)
\label{eq17}
\ee
The formula for $\mu(x)$ is then
\be
\mu(x_0,\vec x) = \exp\left(i \int d^3 y\,\Pi(x_0,\vec y)\bar\varphi(\vec y-\vec x)
\right)
\label{eq18}
\ee
giving
\be
\mu(x) |\varphi(x_0,\vec x)\rangle = |\varphi(x_0,\vec x) + \bar\varphi(\vec x)
\rangle \label{eq19}
\ee
Eq.(\ref{eq19}) is the analog of the familiar translation for a 1 dimension
system
\be
e^{ipa} |x\rangle = |x + a\rangle\label{eq20}
\ee
The construction of $\mu$ for monopoles in compact $U(1)$ gauge theories, in
$SU(N)$ gauge theories\cite{20,24}, in full QCD\cite{24} and for vortices in QCD\cite{14}
has been developed in a series of papers. Similar constructions have been done for
spin system exhibiting duality\cite{5,6,8}. Most of these systems have well known
phase structure, and the construction was made to test the method sketched above,
a test which proved successfull in all cases. We will list here the problems
encountered and solved, and then present the results obtained and discuss their
implications.\\
{\bf 1. Compactness}
In compact theories the fields are
defined in a finite range, and cannot be shifted by arbitrary amount, as in
eq.(\ref{eq19}).

We shall discuss compact $U(1)$, with Wilson's action 
\[ S = \beta\sum_{\mu,\nu,n} (1-\cos\theta_{\mu\nu}(n))\qquad
Z = \exp(-S)\]
The creation operator of a monopole
at $\vec x$ and time $x_0$, can be written in this case
\be
\mu(x) =
\exp\left[\beta\sum_{n,i}\left\{
\cos(\theta_{0i}(\vec n,x_0)-b_i(\vec x-\vec n)) -\cos(\theta_{0i}(\vec n,x_0)
\right\}\right]
\label{eq21}
\ee
For small angles $\theta_{0i},b_i\ll 1$,
$
\mu(x)\simeq\exp\left[
\beta \theta_{0i}(\vec n,x_0) b_i(\vec x - \vec n)\right]$, 
$\theta_{\mu\nu} = \Delta_\mu\theta_\nu - \Delta_\nu\theta_\mu$ is the field
strength tensor, $\theta_{0i} = \vec E_i$ the electric field, i.e.
 $\mu(x)$ reduces to the usual non compact formula (\ref{eq18}) for
small fields. When computing $\langle\mu\rangle = \int [{\cal D}\theta]
\mu \exp(-S)/Z$ it is easily realized that
\begin{equation}
\langle \mu\rangle = {\tilde Z}/{Z}\label{ztilde}
\end{equation}
where $\tilde Z = \exp(-\tilde S)$ and $\tilde S$ is obtained from $S$ replacing
$\cos(\theta_{0i}(\vec n, x_0))$ at time $x_0$ in the action by
$\cos(\theta_{0i}(\vec n, x_0)- b_i(\vec x-\vec n))$.

It is a theorem\cite{10} that $\mu$ creates on the lattice a monopole at $t=x_0$
which then propagates at $t> x_0$. The form (\ref{ztilde}) for the disorder
parameter is common to all systems with duality.\\
{\bf 2. Fluctuations} 
$\langle\mu\rangle = \tilde Z/Z$ is not easy to
compute numerically like any partition function. Indeed $\langle\mu\rangle$ is
(eq(\ref{eq21})) the vev of the exponential of an integral extending to the 3d
space, which is roughly proportional to $N_S^3$, and fluctuates as $N_S^{3/2}$,
producing
for $\langle\mu\rangle$ fluctuations ${\cal O}(\exp( N_S^{3/2})$. The way
out\cite{20} is to compute the quantity
\be
\rho = \frac{d}{d\beta}\ln\langle\mu\rangle = \langle S\rangle_S -
\langle \tilde S\rangle_{\tilde S} \label{eq24}
\ee
\par\noindent
\begin{figure}[htb]
\begin{minipage}{0.5\linewidth}
\epsfxsize\linewidth
\epsfbox{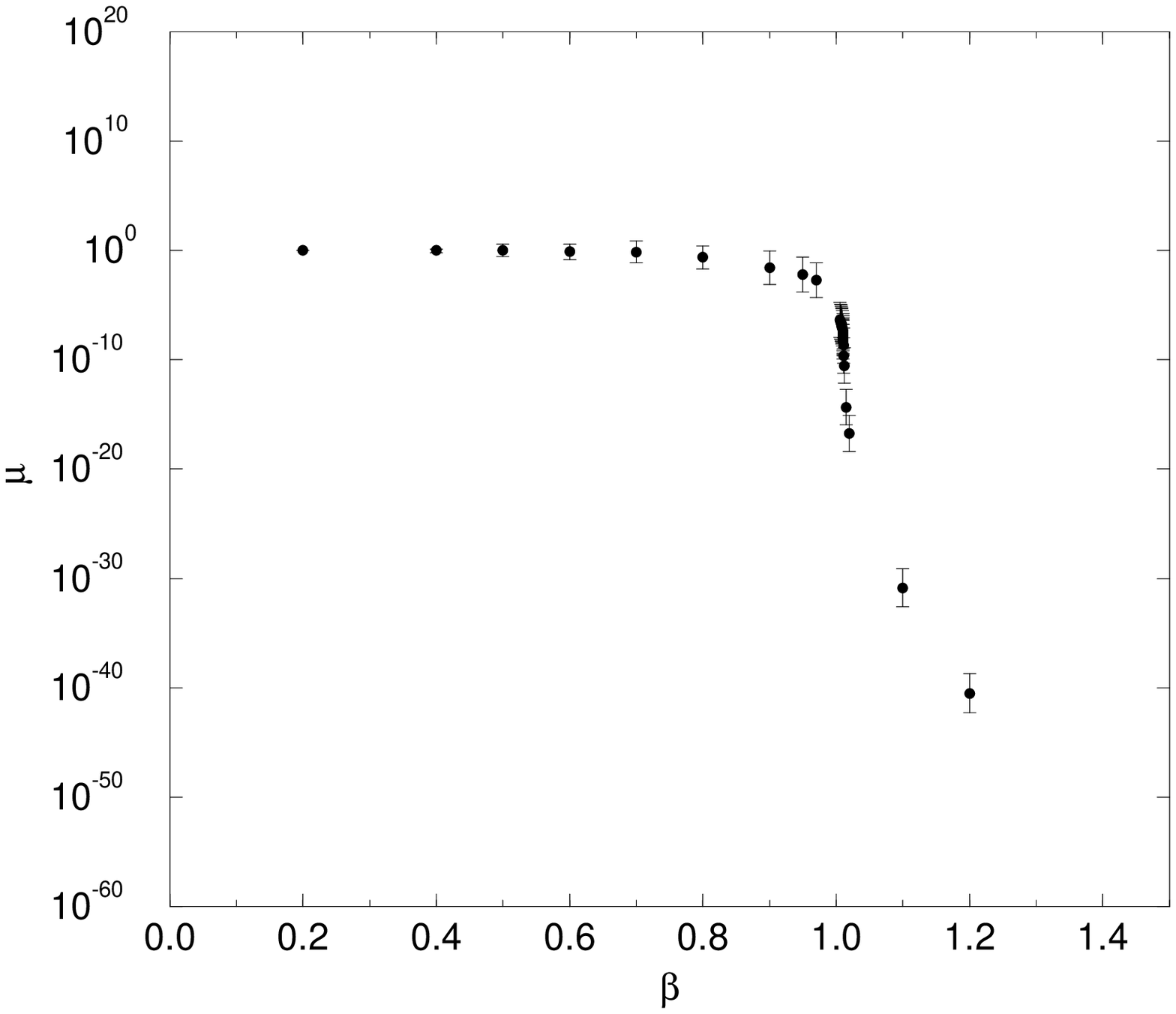}
\caption{Fig.1 $\langle\mu\rangle$ for compact $U(1)$.}
\end{minipage}
\begin{minipage}{0.5\linewidth}
\epsfxsize\linewidth
\epsfbox{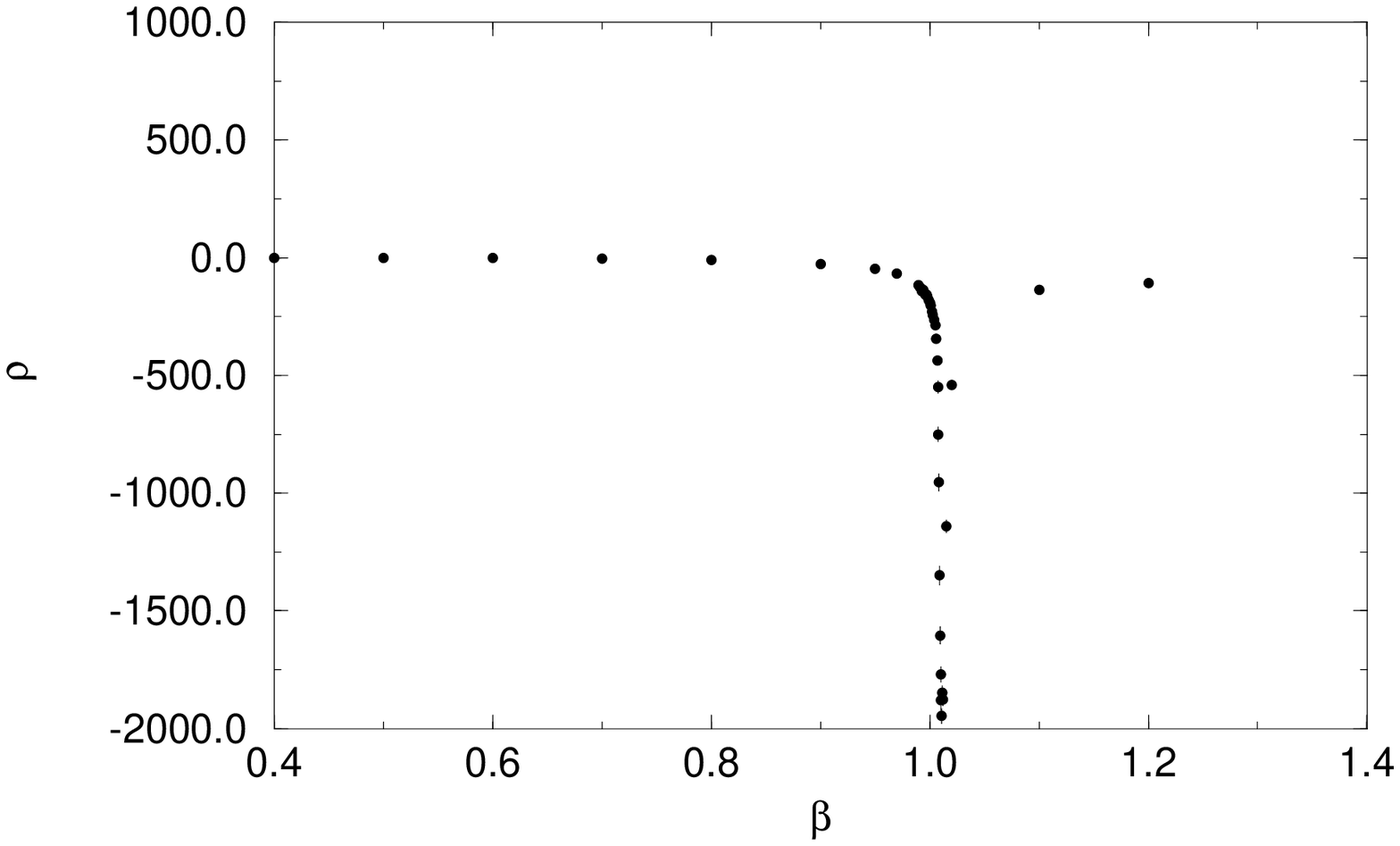}
\caption{Fig.2 $\rho$ for compact $U(1)$}.
\end{minipage}
\end{figure}
which is the difference of the expectation values of two actions. All the relevant
information contained in $\langle\mu\rangle$ can be extracted from $\rho$. The
thermodynamical limit is reached by finite size scaling analysis. A typical shape
of $\langle\mu\rangle$ is shown in fig.(1), which describes the deconfining
transition in compact $U(1)$\cite{10}. $\langle\mu\rangle$ is ${\cal O}(1)$ for
$\beta < \beta_c$ and drops by 50 orders of magnitude at $\beta_c$: in the
infinite volume limit the fall becomes sharper and $\langle\mu\rangle$ is 0
for $T > T_C$.
The corresponding form of $\rho$\cite{10} is shown in fig.(2).

In terms of $\rho$
$
\langle\mu\rangle = \exp\left[\int_0^\beta\rho(x) dx\right]$.
For $\beta < \beta_c$, $\rho$ goes to a finite limit as $V\to\infty$.

For $\beta > \beta_c$, $\rho\to -\infty$, as $V\to\infty$. Indeed $\rho = - k N_S +
k'$, with $k>0$ and $N_S$ the spatial extension of the lattice: this corresponds to
$\lim_{V\to\infty}\langle\mu\rangle = 0$, $\beta >\beta_c$.

Around $\beta_c$
\be
\langle\mu\rangle \mathop\sim_{T\to T_c^-} \left( 1 - \frac{T}{T_C}\right)^\delta
\label{eq26}
\ee
$\delta$ being the critical index of the order parameter.\\
{\bf 3. Locality}
In principle the sum at the exponent in eq.(\ref{eq21}) could produce infrared
problems, e.g. destroying cluster property
\be
\langle\mu(x)\tilde \mu(0)\rangle \simeq C\exp(-A |x|) + |\langle \mu\rangle|^2
\label{eq27}\ee
which is needed to define properly $\langle\mu\rangle$.\\
{\bf 4. Gauge invariance} 
The vev of a (magnetically) charged operator can
have problems with gauge invariance\cite{27}.

All these problems have been solved and checked on the systems where the result is
known, see eg.\cite{10}. 

We will briefly comment on point 3 and 4.
\section{Gauge invariance}
In a gauge invariant formulation of a $U(1)$ gauge theory, the ground state is
gauge invariant ($U_\Lambda |0\rangle = |0\rangle$).  If $\varphi(x)$ is a charged
local operator $U^\dagger_\Lambda\varphi(x) U_\Lambda =
\varphi(x)\exp(i\Lambda(x)$, and hence
\be
\langle0|\varphi(x)|0\rangle = \langle0|U^\dagger_\Lambda\varphi(x)
U_\Lambda|0\rangle = e^{i\Lambda(x)}
\langle0|\varphi(x)|0\rangle \qquad \forall \Lambda
\label{eq27a}
\ee
Therefore $\langle0|\varphi(x)|0\rangle = 0$.

The lattice version of this statement is known as Elitzur's theorem\cite{27}.

The usual way out is to fix the gauge (e.g. the unitary gauge) and look for
spontaneous breaking of $U(1)$ symmetry\cite{30}.
A lattice formulation is per se gauge invariant. How can then a Higgs phenomenon
be described?

The solution consists in defining a gauge invariant charged operator
$\tilde\varphi(x)$, and in taking $\langle0|\tilde\varphi(x)|0\rangle$ as an order
parameter. The definition of $\tilde \varphi$ is the following
\be
\tilde\varphi(x) = \varphi(x)\exp[ i(h,A)]\label{eq28}
\ee
where
\[ (h,A) = \int d^4y A_\mu(y) h^\mu(y-x)\qquad {\rm and}\quad \partial^\mu h_\mu(z)
=
\delta^4(z)\]
Under a gauge transformation $\Lambda(x)$, with $\Lambda(x)\to 0$ as $|x|\to\infty$
$A_\mu\to A_\mu + \partial_\mu\Lambda$ and, by partial integration
$(h,A)\to (h,A) - \Lambda(x)$, while $\varphi(x)\to e^{i\Lambda(x)}\varphi(x)$.
The phase $\Lambda(x)$ cancels between the two factors of eq.(\ref{eq28}) and
$\tilde\varphi(x)\to\tilde\varphi(x)$: $\tilde\varphi(x)$ is gauge invariant.
Under a global transformation instead $A_\mu\to A_\mu$, $(h,A)\to (h,A)$,
$\varphi\to e^{i\Lambda}\varphi$ and
\be
\tilde\varphi\to e^{i\Lambda}\tilde\varphi\label{eq29}
\ee
The exponential factor in eq.(\ref{eq28}) makes $\tilde \varphi$ non local.

A popular choice for $h^\mu$ corresponds to a Schwinger parallel transport. E.g.
if the string is put along the time axis
$h^\mu(z) = \delta^\mu_0 \theta(z_0) \delta^3(\vec z)$ and
\be
\tilde\varphi_S(x) = \varphi(x)\exp\left(i \int_\infty^{x_0} A^0(\vec x,\tau)d\tau
\right) \label{eq30}
\ee
Another possibility is Dirac choice\cite{31}
\[
A^\mu(z) = (0,\delta(z_0) \vec C(\vec z))\qquad
\vec C(\vec z) = \frac{1}{4\pi}\frac{\vec z}{|\vec z|^3}
\] 
for which
\be
\tilde\varphi_D(x) = \varphi(x)\exp\left(i\int d^3 y \vec A(\vec y,x_0)
\frac{1}{4\pi}\frac{\vec y - \vec x}{|\vec y-\vec x|^3}\right)
\label{eq31}
\ee
It can be shown that the correlator
 $\langle\tilde \varphi_S(x)\tilde \varphi_S(0)\rangle$ has infrared problems, in
that it vanishes exponentially at large distances
$ \langle\tilde \varphi_S(x)\tilde \varphi_S(0)\rangle \mathop\to_{|x|\to\infty}
A \exp(-\rho |x|)$
even in the Higgs phase, due to the presence of the string. It is then not suited
to define the order parameter.

The Dirac choice instead is local enough to allow cluster property, and
\be
\langle\tilde \varphi_D(x)\tilde \varphi_D(0)\rangle \simeq
|\langle\tilde\varphi\rangle|^2 +
A \exp(-\rho |x|) \label{eq32}
\ee
with $\langle\tilde\varphi\rangle\neq 0$ in the Higgs phase, $
\langle\tilde\varphi\rangle = 0$ in the Coulomb phase. Eq.(\ref{eq32}) defines
the order parameter.

The operator $\mu$ which creates a monopole\cite{10} is gauge
invariant,magnetically charged, and Dirac like\cite{9,34}.

No infrared problem nor problem with gauge invariance exists.
\par\noindent
\begin{figure}[htb]
\begin{minipage}{0.5\linewidth}
\epsfxsize\linewidth
\epsfbox{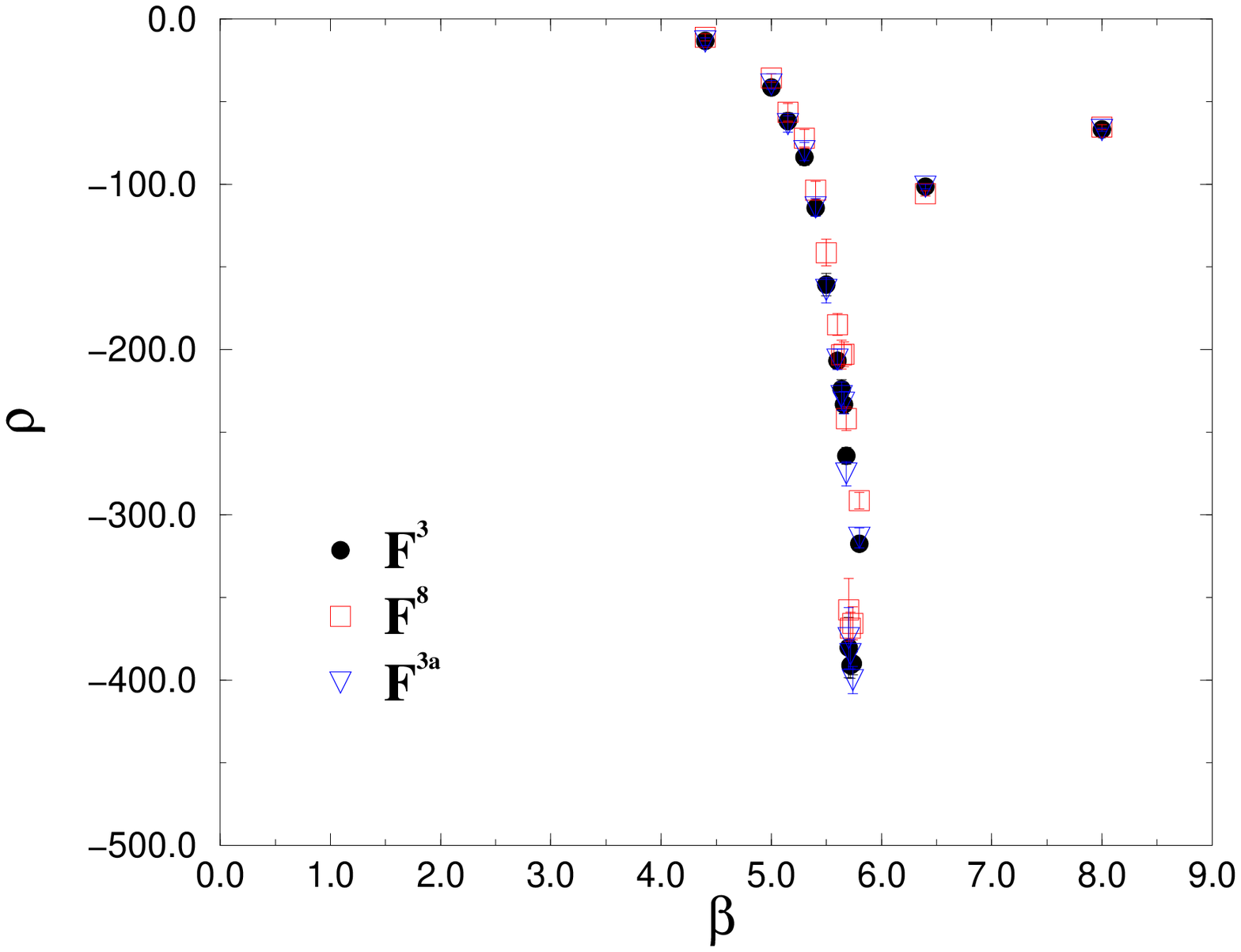}
\caption{Fig.3 $\rho$ for monopole condensation in $SU(3)$.}
\end{minipage}
\begin{minipage}{0.5\linewidth}
\epsfxsize\linewidth
\epsfbox{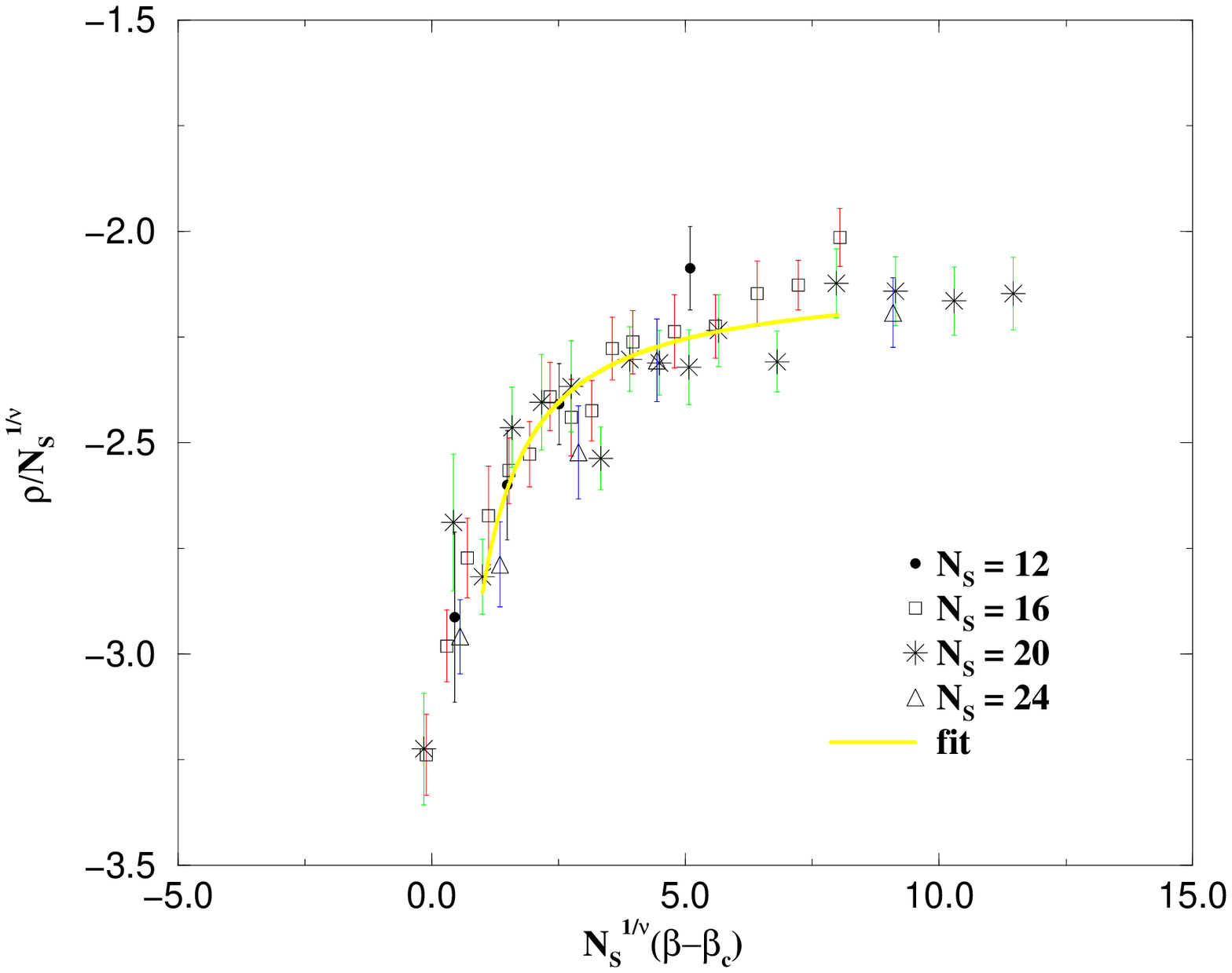}
\caption{Fig.4 Finite size scaling of $\rho$.}
\end{minipage}
\end{figure}
\section{Results for confinement.}
We can summarize the results as follows.
For quenched QCD $\langle\mu\rangle$ can be defined by a recipe analogous to that
of compact $U(1)$\cite{20,24} in any abelian projection and $\rho =
\frac{d}{d\beta}\ln\langle\mu\rangle$ can be determined. The typical behaviour of
$\rho$ is shown in fig.(3) and is qualitatively similar to that in fig.(2) for
$U(1)$.

In order to perform the infinite volume limit $\rho$ is measured for different
spatial sizes of the lattice.
\begin{itemize}
\item[] For $\beta < \beta_c$, $\rho$ goes to a finite limit as $N_S\to \infty$.
\item[] For $\beta > \beta_c$, $\rho\sim -k N_S - k'\quad (k > 0)$, i.e.
$\lim_{N_S\to\infty} \langle\mu\rangle =
\exp(\int_0^\beta \rho(x) dx) = 0 $
\end{itemize}
At $\beta\sim\beta_c$, by dimensional arguments,
\[ \langle\mu\rangle\simeq
(\beta-\beta_c)^\delta\Phi(\frac{N_S}{\xi},\frac{a}{\xi},\frac{N_T}{N_S})\]
where $\xi \sim (\beta-\beta_c)^{-\nu}$ is the correlation length, which goes
large with the index $\nu$ as $\beta\to\beta_c$. Near $\beta_c$ $a/\xi\sim 0$,
$N_T/N_S\sim 0$ and
$\langle\mu\rangle \sim \tau^\delta\Phi(N_S^{1/\delta}\tau)$ or
\[ \frac{\rho}{N_S^{1/\delta}} = \frac{\delta}{\tau} + \Phi(N_S^{1/\delta}\tau)
\]
This scaling law is obeyed, fig.(5), with the known values of $\beta_c$ and $\nu$
and allows to determine $\delta$.

The result for $SU(3)$ is $\delta = 0.50(2)$ independent of the abelian projection.

The computation has been performed
in a number of abelian projections, and also by
choosing\cite{26}, configuration by configuration, as direction of abelian
projected $\Phi$ the conventional 3 axis used to prepare the configuration. This
gives an average on a number of abelian projections equal to the number of
configurations used, and again gives the same behaviour for $\langle\mu\rangle$.

A similar operator $\mu$ can be defined which creates a t'Hooft line. Fig.(6)
shows a comparison between the corresponding $\rho$ and that of monopoles, showing
that they almost overlap.
\section{Full QCD}
In full QCD  $\langle\mu\rangle$ can be defined in the same way as in quenched, by
modifying only the gauge part of the action in the definition of $\tilde S$,
eq.(\ref{eq21}). Simulations give a behaviour similar to the quenched case, and a
negative peak at a $\beta$ which coincides with that of the chiral
transition\cite{32}.

A full finite size scaling analysis is on the way, as well as an analysis of
$\rho$ at $\beta \gg \beta_c$ to establish that $\langle\mu\rangle$ is a correct
disorder parameter.

In principle dual superconductivity does not interphere with the presence of
quarks, so that $\langle\mu\rangle$ would be a good parameter in the line of the
philosophy $N_c\to\infty$.
\begin{center}
\begin{minipage}{0.5\linewidth}
\epsfxsize\linewidth
\epsfbox{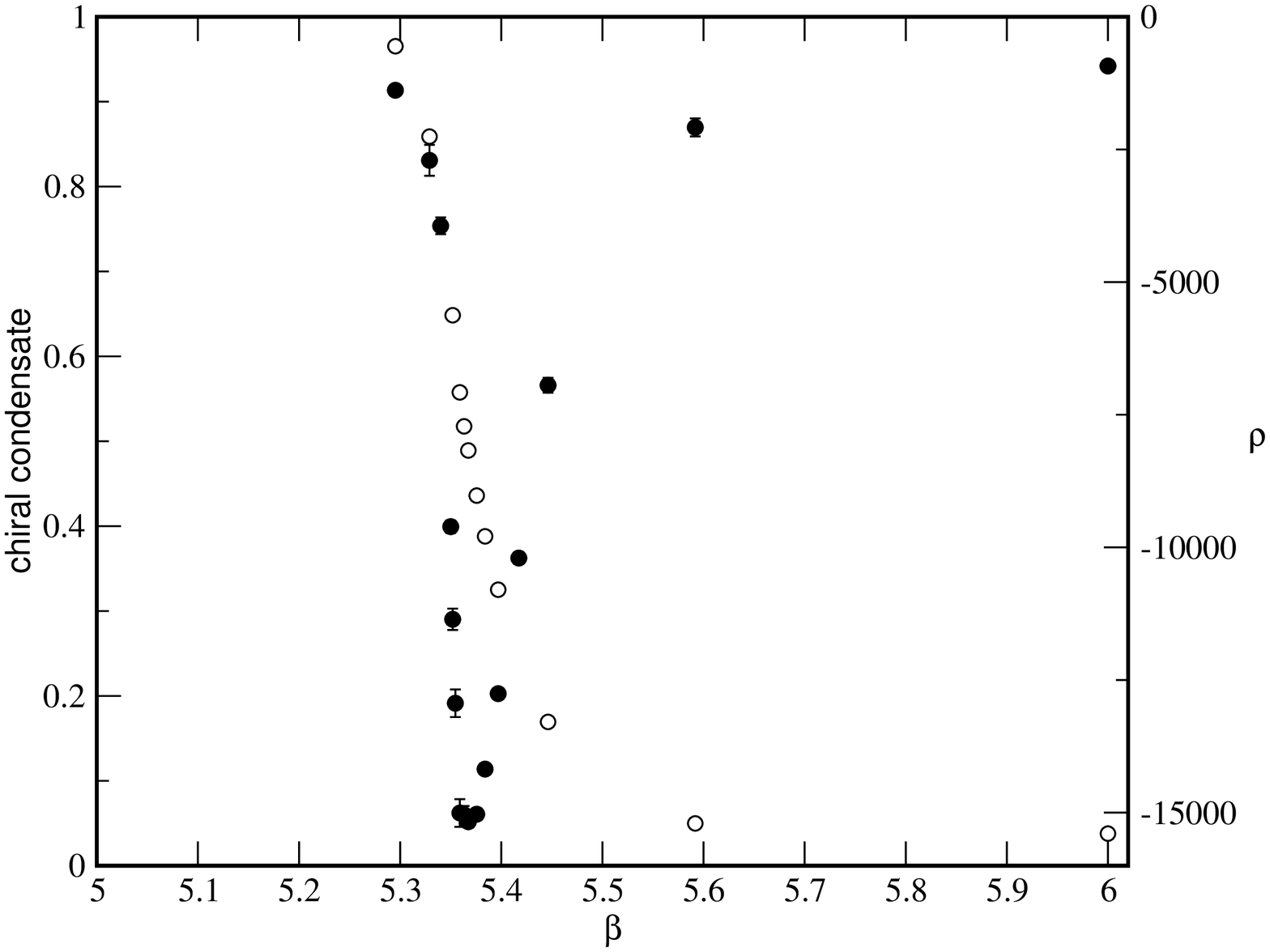}

{Fig.5 $\rho$ for monopole condensation 
(full circles), and chiral condensate (open circles)
for 2 flavour QCD.}
\end{minipage}
\end{center}
\section{Conclusions and outlook.}
\begin{itemize}
\item[1)] deconfinement is an order disorder transition, and the mechanism is dual
superconductivity. Monopoles condense in the confined phase in all abelian
projections, certainly in quenched QCD, and also, by preliminary evidence, in full
QCD, $N_C\to\infty$ limit is respected.
\item[2)]
The dual configurations of QCD are not known. Whatever the dual configurations are,
they carry magnetic charge in all abelian projections.
\item[3)] The dual Polyakov line is a good disorder parameter, and has the same
behaviour as the monopole $\langle\mu\rangle$\cite{14}.
\end{itemize}

The analysis of full QCD must be completed.It will require much computer time and
work.

A breaktrough is needed to identify the dual excitations. A possible direction
could be that indicated in ref.\cite{33}, in which a disorder parmeter is defined
which coincides with ours ${\cal O}(a^2)$ in all abelian projections.

I am grateful to L. Del Debbio, M. D'Elia, B. Lucini, G. Paffuti for discussions
and help in preparing this talk, which is based on our common work.

\end{document}